\documentclass[aps,prb,twocolumn,reprint,superscriptaddress,showkeys,floatfix,amsmath,amssymb]{revtex4-1}

\usepackage{natbib}
\usepackage[T1]{fontenc}
\usepackage{epsfig}
\usepackage[utf8]{inputenc}
\usepackage{mathrsfs}
\usepackage{color}
\usepackage{graphicx}
\usepackage{placeins}
\usepackage{xcolor, soul}
\usepackage{dcolumn}
 \usepackage{hyperref}
\usepackage{bm}
\usepackage{bm}
\usepackage{natbib}
\begin{document}

\preprint{APS/123-QED}

\title{ A first-principles investigation of altermagnetism in CrSb$_2$ under applied pressure}% Force line breaks with \\

	\author{R. Tamang}
	%\author{Zosiamliana Renthlei}
	\affiliation{Department of Physics, Mizoram University, Aizawl-796004, India}%
	\affiliation{Physical Sciences Research Center (PSRC), Department of Physics, Pachhunga University College,  Aizawl-796001, India}
\author{Shivraj Gurung}
	\affiliation{Physical Sciences Research Center (PSRC), Department of Physics, Pachhunga University College,  Aizawl-796001, India}
\author{Shalika Ram Bhandari}
\affiliation{ Department of Physics, Bhairahawa Multiple Campus, Tribhuvan University, Nepal}%

\author{Matthew J. Stitz }
\affiliation{Perry College of Mathematics, Computing, and Sciences, University of West Georgia, Carrollton, GA 30118, USA}
\author{Ganesh Pokharel}
\affiliation{Perry College of Mathematics, Computing, and Sciences, University of West Georgia, Carrollton, GA 30118, USA}
\author{Keshav Shrestha}
\affiliation{Department of Chemistry and Physics, West Texas A\&M University, Canyon, Texas 79016, USA}%
\author{D. P. Rai}
\affiliation{Department of Physics, Mizoram University, Aizawl-796004, India}%
	\email[D. P. Rai]{dibyaprakashrai@gmail.com}
%	\affiliation{Department of Physics, Mizoram University, Aizawl-796004, India}%
	
	\homepage{www.mzu.edu.in}

\date{\today}% It is always \today, today,

\begin{abstract}
In this study, we employed first-principles density functional theory (DFT) calculations within the GGA+U framework to explore the electronic and magnetic properties of CrSb$_2$ under varying hydrostatic pressures. CrSb$_2$ exhibits non-relativistic spin splitting (NRSS) of $\sim$0.5 eV around the Fermi level and the $d$-wave symmetric Fermi surface. Our magnetic susceptibility measurements further confirm the collinear antiferromagnetic (AFM) ground state in CrSb$_2$, a prerequisite for altermagnetism. The presence of collinear AFM and spin-band splitting without the application of spin-orbit coupling (SOC) supports CrSb$_2$ as a potential contender for altermagnet.  With increasing pressure, we have observed an intricate evolution of spin splitting in the valence and conduction bands, governed by changes in orbital contributions. The observation of the structural phase transition above 10 GPa is in qualitative agreement with the previous experimental findings. Our results not only support the classification of CrSb$_2$ as an altermagnetic candidate but also provide critical insight into the role of pressure in tuning its spin-dependent electronic structure.
\end{abstract}

\maketitle

%\tableofcontents
\section{Introduction}
\begin{figure}[hbtp]
	\centering
	\includegraphics[width=1.0\linewidth]{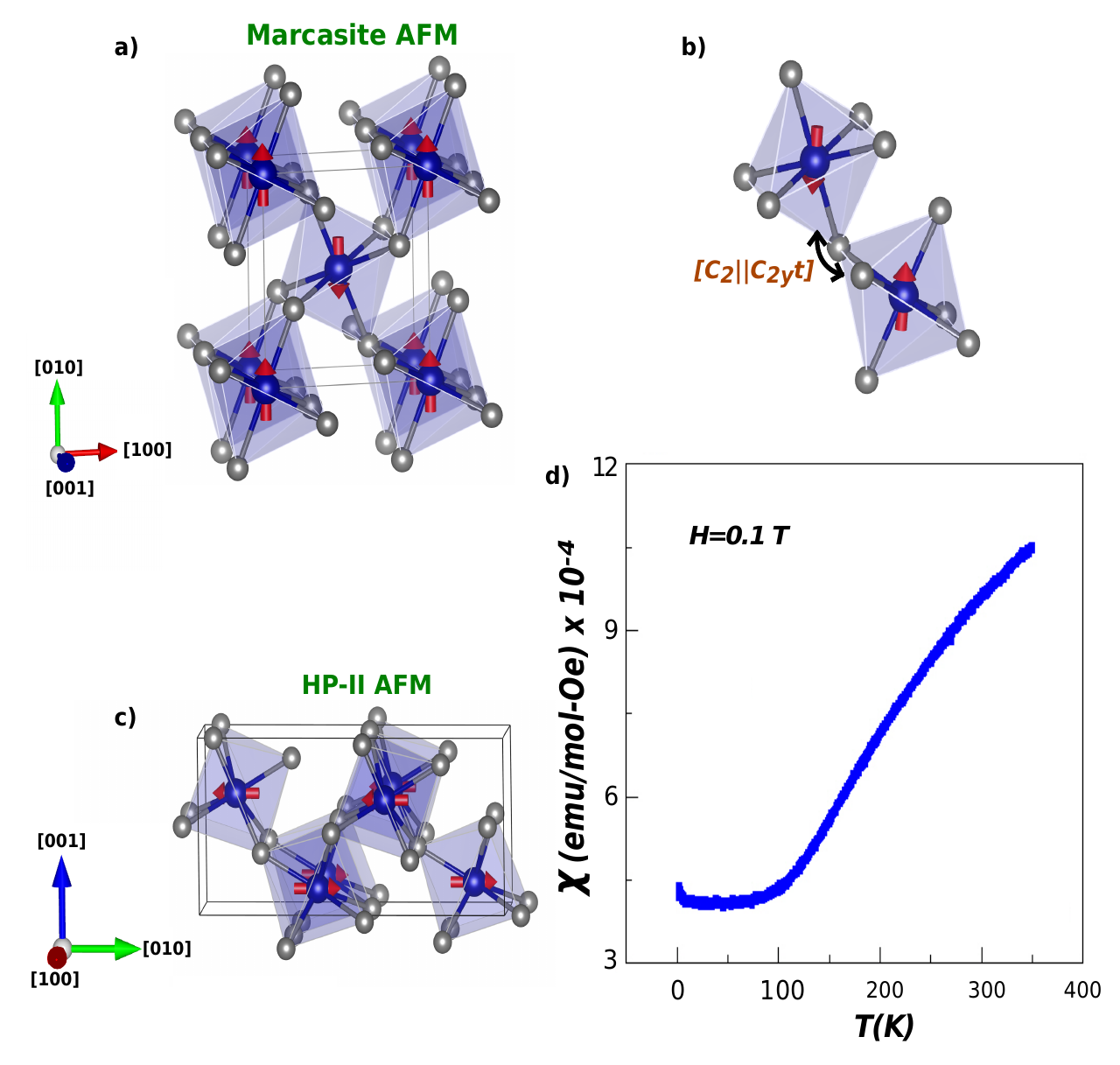} % Adjust width as needed
	\caption{a)The unit cell structure of CrSb$_2$, with magnetic Cr atoms depicted in blue and non-magnetic atoms in grey. b)The opposite spin sublattices are related by a two-fold rotation in both spin space and real space, ([C$_2$$|$C$_{2y}$t]), followed by a half-unit-cell translation. c) HP-II AFM phase of CrSb$_2$. (d) Magnetic susceptibility ($\chi$) as a function of temperature measured under an applied field of $H = 0.1$ T for the marcasite-CrSb$_2$. The susceptibility $\chi(T)$ decreases with decreasing temperature, exhibiting behavior consistent with antiferromagnetism, with a Neel temperature above room temperature.}
	\label{Fig1}
\end{figure}

In recent years, altermagnets (AMs), a newly identified magnetic material, have garnered significant attention in the scientific community, evidenced by the publication of several hundred research articles on the topic. AMs combine essential features of antiferromagnets (AFMs), such as compensated collinear magnetic ordering, and ferromagnets (FMs), including time-reversal symmetry($\mathcal{T}$) breaking and lifting Kramer's degeneracy in the band structures without relativistic spin-orbit coupling (SOC). However, they exhibit distinct symmetries that separate them from both AFM and FM \cite{vsmejkal2022emerging,vsmejkal2022beyond}.  In traditional collinear AFMs, the symmetry that connects the two opposite spin sublattices is $\mathcal{T}$ combined with parity($P$) or transition($t$). However, in altermagnets (AMs), due to the specific crystal environment around each magnetic atom, the opposite spin sublattices are related by rotations, screws, glide, and mirrors \cite{song2025altermagnets} the opposite spin sublattices are related by rotations in both spin and real space followed by translation [see Fig.\ref{Fig1}(b)].
 While FMs are prone to stray fields, which can cause device malfunctions \cite{baltz2018antiferromagnetic}. On the other hand, AFMs lack spin polarization. In this context, AMs emerge by leveraging the advantages of both conventional magnetic phases, FMs and AFMs, while exhibiting non-relativistic spin splitting (NRSS) in their band structure, concurrently achieving immunity to stray magnetic fields. \cite{vsmejkal2022emerging,vsmejkal2022beyond,mazin2022altermagnetism}. NRSS is comparable to the spin-splitting induced by relativistic SOC on the order of a few electron volts\cite{vsmejkal2022emerging,vsmejkal2022beyond}.

 First-principles calculations have identified several bulk and two-dimensional materials as potential altermagnets with distinct NRSS\cite{bai2024altermagnetism,guo2023spin,vsmejkal2022emerging}. However, only a few materials have been experimentally confirmed to exhibit NRSS, as investigated through angle-resolved photoemission spectroscopy (ARPES)\cite{zhu2024observation,lee2024broken,ding2024large,reimers2024direct,yang2025three}. Various spin-dependent phenomena can emerge in this novel magnetic phase. Notably, the anomalous Hall effect (AHE), traditionally considered forbidden in collinear antiferromagnets due to the preservation of time-reversal symmetry ($\mathcal{T}$), has recently garnered significant attention. Anomalous Hall conductivity represents a dissipationless portion of the conductivity tensor\cite{vsmejkal2020crystal}. Contrary to conventional understanding, theoretical predictions indicated the possibility of AHE in collinear antiferromagnetic RuO$_2$, a finding that has been experimentally validated in altermagnetic RuO$_2$\cite{feng2022anomalous,tschirner2023saturation}. Subsequent experimental confirmations of AHE have been reported in other altermagnetic materials, including Mn$_5$Si$_3$ \cite{reichlova2024observation,leiviska2024anisotropy}, MnTe\cite{kluczyk2024coexistence,gonzalez2023spontaneous}, and CrSb \cite{zhou2025manipulation}. These observations collectively confirm the spontaneous breaking of $\mathcal{T}$-symmetry in AMs. Additionally, altermagnets (AMs) exhibit significant potential for spintronic applications, owing to the generation of spin currents \cite{shao2021spin,gonzalez2021efficient,bose2022tilted,naka2021perovskite,naka2019spin}, giant and tunnel magnetoresistance effects\cite{vsmejkal2022giant, jiang2023prediction,chi2024crystal,das2023transport,liu2024giant}, spin-splitter torques\cite{bai2022observation,karube2022observation}, efficient spin-to-charge conversion\cite{bai2023efficient,zhang2024simultaneous}, and ultrafast magnetic dynamics operating in the terahertz (THz) regime\cite{baltz2024emerging,vsmejkal2023chiral}. These phenomena position AMs as highly promising candidates for next-generation spintronic technologies\cite{song2025altermagnets}.

 \begin{figure}[hbtp]
	\centering
	\includegraphics[width=.75\linewidth]{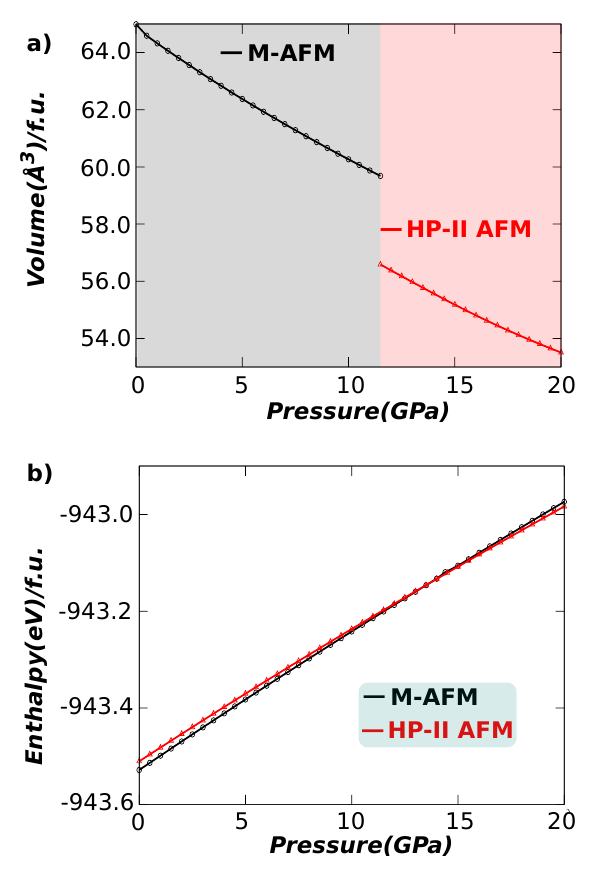} % Adjust width as needed
	\caption{a) Variation of volume per formula unit(f.u.) with applied pressure in both marcasite(M) and HP-II phase of antiferromagnetic CrSb$_2$. b) Enthalpy per formula unit(f.u.) vs. pressure curve with crossover point above 10 GPa. The HP-II phase is thermodynamically more favourable above 10 GPa.}
	\label{EV}
\end{figure}
\begin{figure*}
\centering
\includegraphics[width=1.00\linewidth]{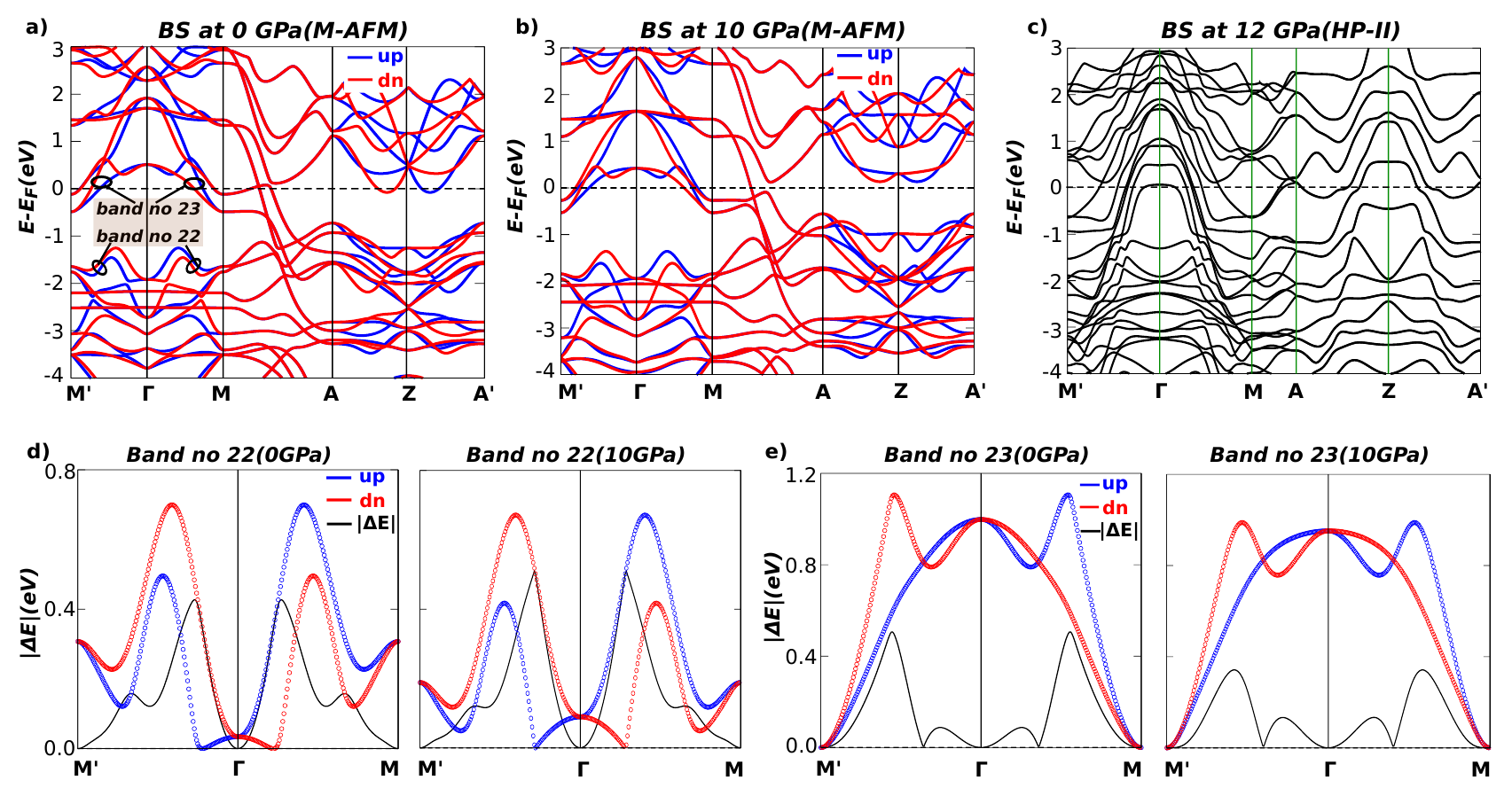}
\caption{Illustration of the Electronic Properties at 0, 10, and 12 GPa: (a-c) marcasite(M) AFM AT 0 GPa, M-AFM at 10 GPa, and HP-II AFM at 12 GPa, respectively. d) Amplification of the NRSS, denoted by $\lvert \Delta E \rvert$, is observed in band index 22 over the pressure range of 0 to 10 GPa. e) In contrast, a significant suppression of   NRSS $\lvert \Delta E \rvert$ is evident in band index 23 within the same pressure interval. }
\label{3}
\end{figure*}

\begin{figure*}
\centering
\includegraphics[width=0.9\linewidth]{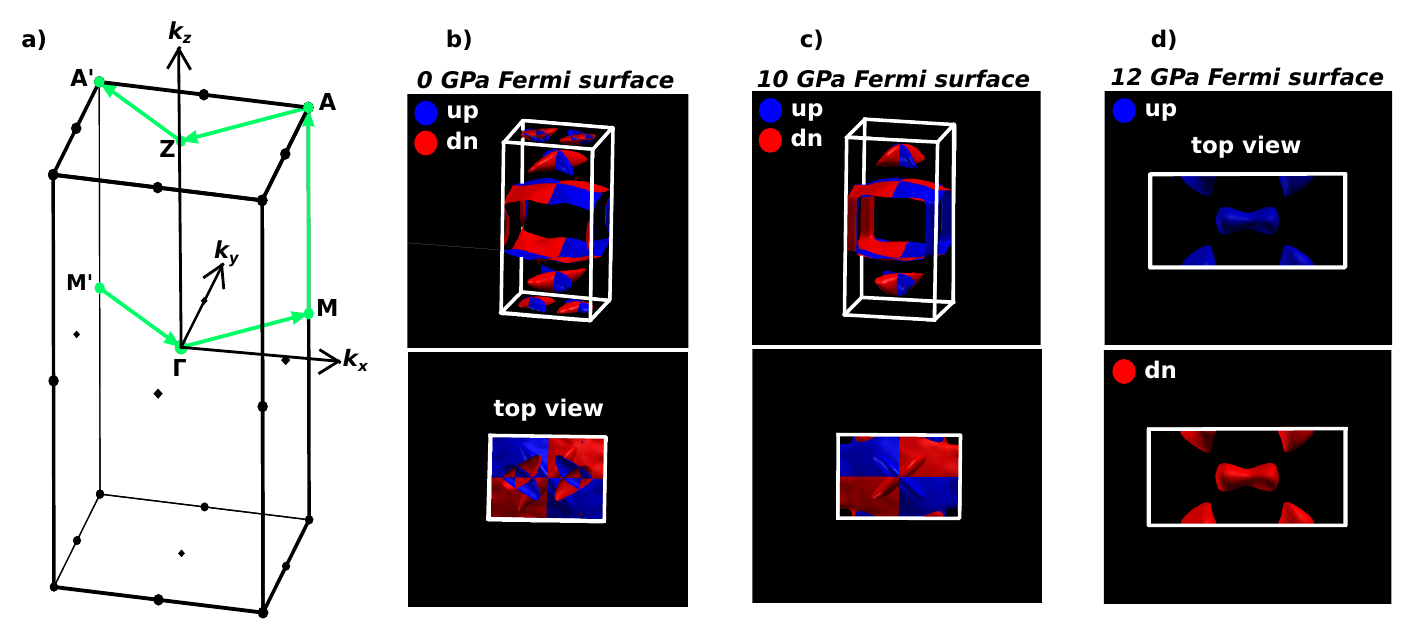}
\caption{Fermi surfaces under varying pressures: a) High-symmetry K-path. (b–c) Fermi surfaces at 0 and 10 GPa, respectively, depicting spin–momentum locking with characteristic d-wave symmetry in the marcasite phase. (d) Fermi surface at 12 GPa in the HP-II phase, showing symmetric spin-up and spin-down channels, confirming the absence of NRSS.}
\label{Fermmi}
\end{figure*}

\begin{figure*}
\centering
\includegraphics[width=1.04\linewidth]{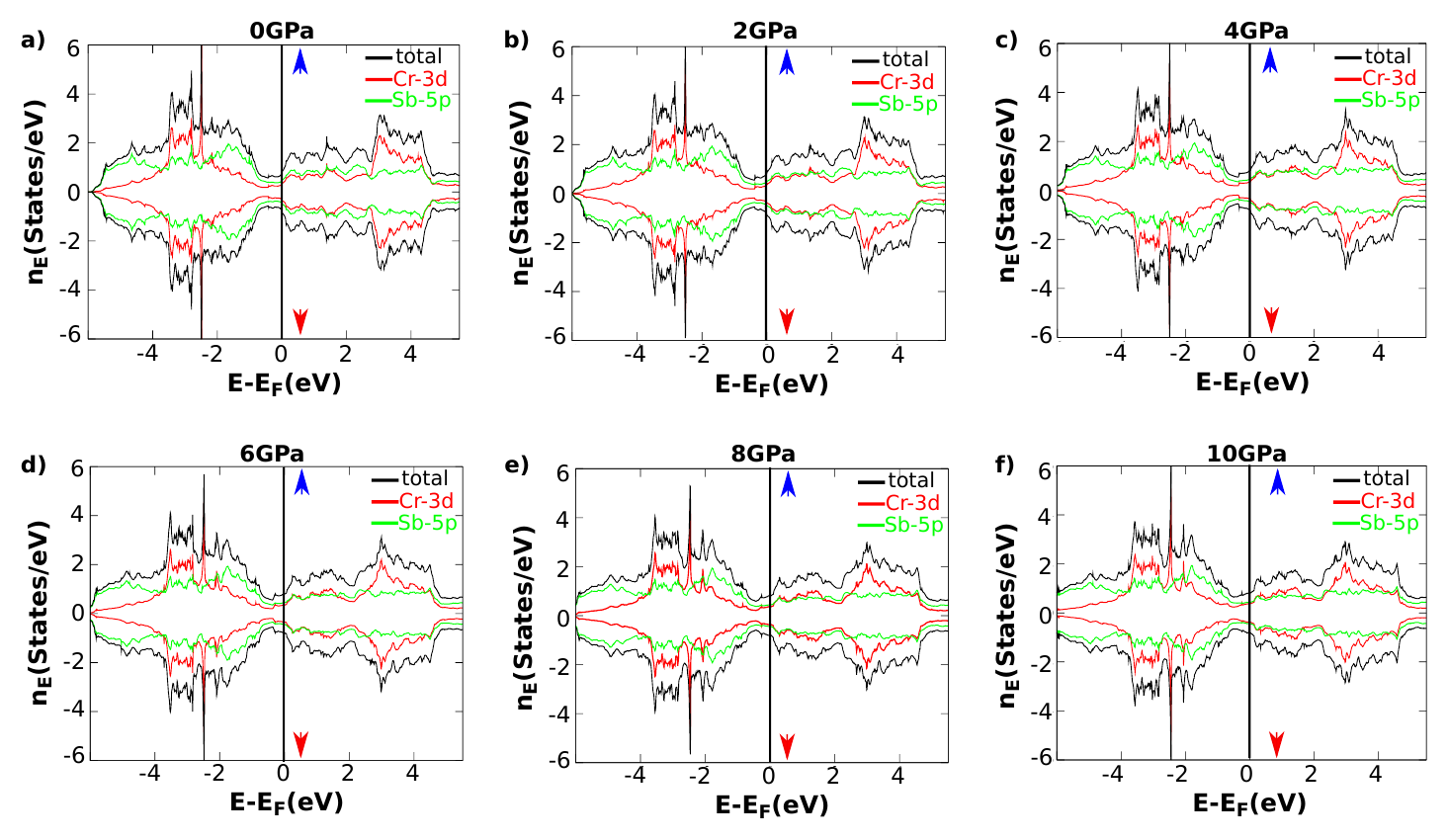}
\caption{Projected Density of States (PDOS) Under Varying Pressure Conditions: a)-f) Spin-resolved PDOS at 0, 2, 6, 8, and 10 GPa, respectively, demonstrating the progressive modification of the projected density of states under applied pressure.}
\label{pdos}
\end{figure*}
 
\begin{figure}[htbp]
	\centering
	\includegraphics[width=1.0\linewidth]{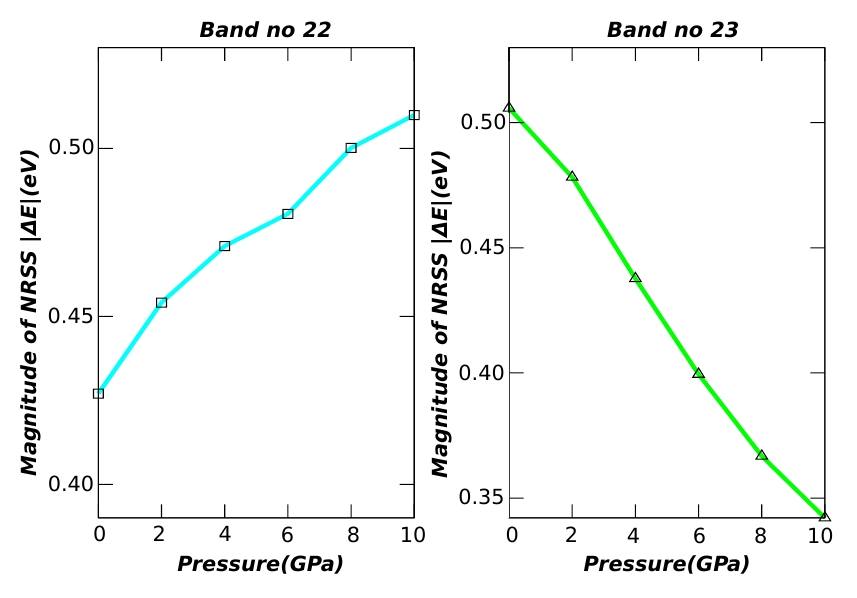} % Adjust width as needed
	\caption{ Variation of spin splitting under applied pressure in the range of 0 to 10 GPa for band indices 22 and 23.}
	\label{splitting}
\end{figure}

Various theories have been developed to describe AM, with one prominent theory utilizing the spin group theory developed by Litvin et al.\cite{litvin1974spin,litvin1977spin}, which has been applied to explain the NRSS in AM.\cite{vsmejkal2022beyond}. Radaelli and group used the tensorial approach to discuss altermagnetism\cite{radaelli2024tensorial}. Bhowal et al.\cite{bhowal2024ferroically} investigated the centrosymmetric antiferromagnet MnF$_2$, emphasizing the role of higher-order multipoles in driving NRSS. Magnetic octupoles in MnF$_2$ break time-reversal symmetry and induce NRSS without relying on relativistic SOC. The spin-splitting energy depends on structural and magnetic modifications, exhibiting a strong correlation with ferroic octupolar ordering. The magnitude of spin splitting can be represented by an equation\\

$E(\mathbf{k}, \uparrow) - E(\mathbf{k}, \downarrow) = F(k) \sin(k_x a) \sin(k_y a)$\cite{mazin2021prediction}\\

Here, $F(k)$ is a k-dependent function, whereas $a$ is the lattice constant. At high-symmetry points in the Brillouin zone, where $k_x=n\pi/a$ and $k_y=n\pi/a$ (n=0,1,..), the spin splitting vanishes, resulting in degenerate band structures along high-symmetry paths.\\

Although CrSb$_2$ has been predicted AM candidate\cite{bai2024altermagnetism} but there are no solid works on the presence of altermagnetism in CrSb$_2$. Moreover, altermagnetism has already been reported in its close cousins like CrSb \cite{reimers2024direct,ding2024large,yang2024three} and FeSb$_2$ \cite{phillips2025electronic}. The experimental studies have confirmed the marcasite AFM structure of CrSb$_2$ at ambient pressure\cite{holseth1970compounds}, analogous to that of FeSb$_2$, along with the observation of comparable Seebeck coefficients \cite{sales2012transport}. Experimentally, the existence of altermagnetism in FeSb$_2$ has already been confirmed via Fermi surface mapping \cite{phillips2025electronic}. Complementary to this, the surface properties of CrSb$_2$ have also been extensively investigated\cite{nakagawa2023surface,du2020surface} and reported a pronounced surface anisotropic magnetoresistance based on angle-dependent magnetoresistance measurements. Furthermore, Mazin and colleagues have reported the presence of an anti-Kramers (AK) nodal surface alongside notable NRSS in Cr doped FeSb$_2$; such an AK nodal surface gives rise to a significant anomalous Hall conductivity ($\sim$150 S/cm), as demonstrated through first-principles calculations\cite{mazin2021prediction}.

We have successfully grown high-quality single crystals of CrSb$_2$ using a conventional flux-growth method. The temperature-dependent magnetic susceptibility ($\chi$) calculation of CrSb$_2$ revealed an AFM ground state. Motivated by the experimental results, we intend to perform the first-principles DFT calculation to explore the possible occurrence of altermagnetism in CrSb$_2$. First-principles calculations based on density functional theory (DFT) provide a fundamental framework for probing the structural, electronic, and magnetic characteristics of materials. These computational approaches are indispensable for elucidating underlying microscopic mechanisms, rationalizing experimental findings, and systematically examining material responses to external perturbations such as pressure or strain\cite{devaraj2024interplay,fan2025high,chakraborty2024strain}. They play a pivotal role in accelerating the rational design and discovery of novel functional materials. Furthermore, by substantially minimizing reliance on iterative experimental trial-and-error, these methods markedly enhance the efficiency and precision of materials research. Notably, the discovery and conceptual development of altermagnetism were initially driven by insights gained through first-principles computations\cite{vsmejkal2022beyond,vsmejkal2022emerging}. Consequently, we analyze how hydrostatic pressure influences the electronic and magnetic properties of the system at 0K, uncovering a strong coupling between pressure-induced structural phase transitions and the onset of altermagnetic behaviour. Our findings highlight the role of lattice compression in modulating spin-dependent band structures. 
 A comprehensive first-principles investigation by Kuhn et al. \cite{kuhn2013electronic} examined various magnetic configurations, including nonmagnetic, ferromagnetic, and antiferromagnetic phases, to elucidate the material's electronic and magnetic ground state. The incorporation of electron correlation effects through the GGA+U approach yields results that align well with experimental findings, precisely capturing the antiferromagnetic ground state of CrSb$_2$. 

\section{METHODOLOGY}
\subsection{Sample Synthesis and Details}
High-quality single crystals of CrSb$_2$ were grown using a conventional flux-growth technique. A self-flux of antimony (Sb) was employed to create a liquid environment conducive to initiating the solid-state reaction at high temperatures. A mixture of chromium (Cr, pieces, 99.99\%) and antimony (Sb, shot, 99.999\% ), both purchased from Alfa Aesar, was prepared in a molar ratio of 1:12 and loaded into a Canfield crucible set. The crucible was then sealed inside a quartz tube under an argon atmosphere at approximately 1/3 atm. 
The sealed ampoule was heated at a rate of 200 $^o$C/h to 1000 $^o$C and held at this temperature for 36 hours to ensure proper homogenization. It was then slowly cooled to 640 $^o$C at a rate of 2 $^o$C/h. At 640 $^o$C, shiny CrSb$_2$ single crystals with typical dimensions of approximately 2 $\times$ 2 $\times$ 1 mm$^3$ were separated from the remaining Sb flux using a centrifuge technique. The structural quality of the crystals was verified using powder X-ray diffraction (XRD) of ground crystals, and their chemical composition was examined via energy-dispersive X-ray spectroscopy (EDS) on single crystals. Magnetization measurements on the CrSb$_2$ single crystal were performed using a Magnetic Property Measurement System (MPMS, Quantum Design).

\subsection{Computational Details}
\hspace{0.5cm}We have performed plane wave-based density functional theory (DFT) calculations using the open-source software package Quantum ESPRESSO\cite{giannozzi2009quantum,giannozzi2017advanced,giannozzi2020quantum} (QE) to investigate the electronic and magnetic properties of CrSb$_2$. The unit cell structure, obtained from the online database The Materials Project, was optimized using the vc-relax approach. The antiferromagnetic ground state was achieved using a fine Monkhorst-Pack k-point mesh of 16$\times$16$\times$16. The marcasite structure of CrSb$_2$ with motif-pair anisotropy and AFM coupling between two opposite spin Cr atoms is shown in Fig.\ref{Fig1}(a-c). To investigate the NRSS in CrSb$_2$, we used ultrasoft non-relativistic pseudopotentials from the PSlibrary and PBEsol exchange correlation function\cite{perdew2007generalized}. Furthermore, the total energy convergence was achieved with a threshold of $10^{-8}$ Rydberg. Within the GGA+$U$ formalism, an on-site Hubbard $U = 2.7$ eV was assumed for the Cr 3$d$ orbitals to account for correlation effects\cite{kuhn2013electronic}. We used energy cutoffs of 70 Ry and 750 Ry for the wave function and charge density, respectively. Hydrostatic pressure, as implemented in QE, was sequentially applied in the range of 0 to 20 GPa. To study the dynamical stability of an optimised structure 2$\times$2$\times$2  q-point was set for phonon dispersion calculations.

\section{Discussion}
Figure~\ref{Fig1}(d) shows the temperature dependence of magnetic susceptibility ($\chi$) for a CrSb$_2$ single crystal measured with an applied magnetic field of $H = 0.1$ T along the c-axis. The $\chi(T)$ curve decreases as the temperature is lowered, strongly indicating an antiferromagnetic ground state below 350 K. The Neel temperature is therefore expected to be above 350 K, beyond the experimental temperature range. Fig.\ref{Fig1}a illustrates the marcasite phase of CrSb$_2$, which crystallizes in the orthorhombic space group $Pnnm$ (No. 58), comprises six atoms per unit cell with fully compensated magnetic moments. In this structure, Cr atoms occupy the 2a Wyckoff position, whereas Sb atoms reside at the 4g Wyckoff site\cite{ehrenreich2023seven}. The Sb octahedra surrounding the magnetic Cr atoms are structurally inequivalent to those surrounding the opposite-spin Cr atoms. This asymmetry gives rise to a screw symmetry operation that connects the two opposite spin octahedra, favoring the formation of altermagnetic order.\\

 Under high-pressure conditions, a notable suppression of the unit cell volume is observed as shown in Fig.\ref{EV}a). This trend aligns well with the first-principles and experimental results\cite{li2023rewritable,ehrenreich2023seven}. At 0 GPa, the optimized lattice parameters of marcasite phase are found to be a=6.026\AA, b=6.846\AA, and c=3.151 \AA, which corresponds to a unit cell volume of 129.991 \AA$^3$. This computed volume shows good agreement with the experimental value of 130.7 \AA$^3$, reported in Ref.\cite{li2023rewritable}. The application of pressure modulates the lattice environment and leads to notable changes in the lattice parameters of CrSb$_2$. The variations of lattice parameters under the investigated pressure range (0–10 GPa) are depicted in the supplementary material, Figure S1.  As pressure increases, CrSb$_2$ undergoes two sequential structural phase transitions. The first occurs above 5.5 GPa, where the system transforms into a CuAl$_2$-type structure (HP-I)\cite{takizawa1999new}. Beyond 10 GPa, a further transition leads to a MoP$_2$-type phase (HP-II)\cite{ehrenreich2023seven}.  The transition is substantiated by first-principles calculations, wherein the enthalpy-pressure curves exhibit well-defined crossover points above 10 GPa at the transition pressures ( illustrated in Fig. \ref{EV}b). The computed phase stability trends demonstrate good agreement with experimental findings, thereby reinforcing the reliability of the theoretical approach in capturing the structural evolution of CrSb$_2$ under a high-pressure regime. At approximately 12 GPa, the optimized lattice parameters for the HP-II phase are $a = 2.978$ \AA, $b = 12.617$ \AA, and $c = 6.002$ \AA, yielding axial ratios of $a/b = 0.236$ and $a/c = 0.496$. These values closely match experimental data for the MoP$_2$\cite{ehrenreich2023seven}. Due to its ferromagnetic ordering, the HP-I phase of CrSb$_2$ was excluded from this study \cite{takizawa1999new}.. Additionally, the electronic properties also evolve under applied pressure (see Fig. \ref{3}). The electronic band structure reveals a maximum spin splitting $E(k,\uparrow)-E(k,\downarrow)$ of approximately 0.5 eV along the M$^{\prime}-\Gamma$$-$M path, near the Fermi level (shown in Fig. \ref{3}e) at 0 GPa). First-principles computations without the inclusion of relativistic effects underscore the potential of CrSb$_2$ as a candidate AM. Notably, our results demonstrate the emergence of alternating spin-momentum locking exhibiting $d$-wave symmetry, as illustrated in Fig. \ref{Fermmi}(b-c). The presence of such anti-Kramers nodal surfaces and nontrivial electronic states is indicative of the generation of a finite anomalous Hall conductivity, even in the absence of net magnetization\cite{mazin2021prediction}. However, the HP-II phase of CrSb$_2$ exhibits no spin–momentum locking, as the spin-up and spin-down channel Fermi surfaces are completely symmetric (as shown in Fig.\ref{Fermmi}d), consistent with the absence of NRSS in the HP-II phase. Fig.\ref{pdos} illustrates the projected density of states (PDOS), showing that the spin-up and spin-down contributions are equal and opposite across the energy spectrum, thereby confirming the persistence of compensated magnetization across all investigated pressure ranges. The electronic band structure of the HP-II phase reveals a definitive absence of NRSS, as depicted in Fig.\ref{3}c, aligning with the findings reported by Linnemann $et$ $al.$\cite{linnemann2024weyl}. Moreover, a large number of band crossings near the Fermi level of the HP-II phase (see Fig.\ref{3}c) can be attributed to the presence of Weyl points, as previously reported in Ref. \cite{linnemann2024weyl}.\\

Alongside modifications in structural and electronic properties, pressure exerts a notable influence on magnetic properties. The magnetic moment ($\mu$) exhibits a decreasing trend with increasing pressure, as illustrated in Table\ref{2}, aligning well with experimental implication\cite{ehrenreich2023seven}. Furthermore, pressure enhances the strength of spin splitting in the valence band, whereas in the conduction band, pressure suppresses the spin splitting( variation of NRSS is illustrated in Fig.~\ref{splitting}). This trend is consistent with the findings of Devaraj $et$ $al.$\cite{devaraj2024interplay}. Such a trend can be rationalized by examining the projected density of states (PDOS). With increasing pressure, interatomic distances are reduced, which enhances the hybridization between p- and d-orbitals. As a consequence, the contribution of d-orbitals to the conduction band slowly decreases, as shown in Fig.~\ref{pdos} \cite{devaraj2024interplay}. This decrement in d-orbital character leads to a suppression of non-relativistic spin splitting (NRSS) in the conduction band\cite{fan2025high}. In contrast, the valence band exhibits the opposite trend: elevated pressure gradually enhances the d-orbital contribution to the valence states, thereby resulting in an enhancement of NRSS.    
 This contrasting behavior highlights the orbital-selective nature of pressure-induced spin splitting and the pivotal role of orbital hybridization in shaping the electronic structure. Furthermore, microscopic parameters are critical in governing the non-relativistic spin splitting (NRSS). In particular, the competition between antiferromagnetic exchange interactions and d-orbital hopping integrals plays a decisive role in modulating NRSS. This relationship can be qualitatively expressed as $\left| \Delta E \right| \propto \frac{t_3 t_4}{J}$\cite{bhowal2024ferroically}
, where $t_3$ and $t_4$ correspond to intra-orbital and inter-orbital hopping parameters of the d-orbitals, respectively(within two nearest sublattices), and $J$ denotes the antiferromagnetic exchange energy. Increasing external pressure tends to suppress the magnitude of $J$, thereby facilitating enhanced electron hopping between orbitals. This results in an effective amplification of the spin splitting \cite{fan2025high}.
 
\begin{table}[htb]
\caption{\label{2}%
Lattice parameters, unit cell volume (V), magnetic moment per Cr atom ($\mu_{Cr}$), and enthalpy per formula unit (E(eV)/f.u.) for antiferromagnetic CrSb$_2$ in the marcasite (M) phase under various applied pressures.}
\begin{ruledtabular}
\begin{tabular}{ccccccc}
P(GPa) & a(\AA) & b(\AA) & c(\AA) & V(\AA$^3$) & $\mu_{Cr}$ ($\mu_B$)& E(eV)/f.u. \\
\hline
0 & 6.026 & 6.846 & 3.151 & 129.991  &2.291 &-943.528  \\
5 & 5.951 & 6.757 & 3.103 & 124.774 & 2.144 & -943.383 \\
10 & 5.894   & 6.679   & 3.062   & 120.539 & 2.039  & -943.242      \\
15 & 5.843   & 6.609 & 3.026    & 116.853 & 1.959   & -943.106      \\
20 & 5.797    & 6.544  & 2.994    & 113.579 & 1.892       & -942.974      \\
\end{tabular}
\end{ruledtabular}
\end{table}

\begin{table}[htb]
\caption{\label{4}%
Lattice parameters, unit cell volume (V), magnetic moment per Cr atom ($\mu_{Cr}$), and enthalpy per formula unit (E(eV)/f.u.) for antiferromagnetic CrSb$_2$ in the HP-II phase under various applied pressures.}
\begin{ruledtabular}
\begin{tabular}{ccccccc}
P(GPa) & a(\AA) & b(\AA) & c(\AA) & V(\AA$^3$) & $\mu_{Cr}$ ($\mu_B$)& E(eV)/f.u. \\
\hline
0 & 3.0965 & 13.056 & 6.164 & 249.198  &2.678 &-943.510  \\
5 & 3.039 & 12.882 & 6.028 & 235.987 & 2.309 & -943.371 \\
10 & 2.992   & 12.689   & 6.028   & 228.856 & 1.928  & -943.237      \\
15 & 2.946   & 12.516 & 5.959    & 219.721 & 1.555   & -943.108      \\
20 & 2.935    & 12.378  & 5.890    & 213.980 & 1.248      & -942.983      \\
\end{tabular}
\end{ruledtabular}
\end{table}
 
 The NRSS can be interpreted using the richer non-relativistic spin group theory, where transformations in spin space and real space can be resolved. This implies that transformations in real space and spin space can act independently. Traditionally, the spin groups are represented by the direct product r$_s$$\times$R$_s$\cite{litvin1974spin,litvin1977spin}, where r$_s$ represents transformations in the spin space alone, and R$_s$(nontrivial spin group) represents a combination of transformations in both real and spin spaces. Three types of nontrivial spin groups describe three magnetic systems, i.e., FM, AFM, and AM\cite{vsmejkal2022beyond}. The spin-only group in the case of collinear spins is given by  $r_s$=$C_{\infty}$+$\bar{C}_2$$C_{\infty}$\cite{litvin1974spin,litvin1977spin}. $C_{\infty}$ represents the group containing all the rotational transformations around the common axis of spin. Meanwhile, $\bar{C}_2$ represents a two-fold rotation around the axis orthogonal to the spin, followed by spin space inversion. The non-trivial spin Laue group that describes the AMs($R^{III}_s$) is constructed using isomorphism theorm\cite{litvin1974spin}.\\
 
 $R^{III}_s=[E||H]+[C_2||G-H]$\cite{vsmejkal2022beyond}\\
 
Here, transformations to the left of the double vertical bars act on spin space, while those to the right act on real space. $G$ denotes the crystallographic Laue group, while $H$ represents its halving subgroup, comprising real-space symmetry operations that map atoms within the same spin sublattice. In contrast, the coset $G-H=AH$ consists of symmetry operations that interchange atoms between opposite spin sublattices \cite{vsmejkal2022beyond}. The operator $A$ represents proper or improper, symmorphic or nonsymmorphic rotations.  However, unlike in the case of AFMs, the two opposite spin sublattices are not related by translation or inversion symmetry. Now $[C_2||AH]\mathcal{E}(\sigma,k)=\mathcal{E}(-\sigma,k')$, thereby giving rise to a spin-split band structure \cite{vsmejkal2022beyond}. For instance, non non-trivial spin Laue group $R^{III}_s$ that characterizes altermagnetism in CrSb$_2$ is\\

 $^2m^2m^1m$=$[E||2/m]$+$[C_2||C_{2y}][E||2/m]$\\
 
 The opposite spin sublattices are related by a non-symmorphic rotation, specifically a twofold rotation (C$_2$) in spin space around an axis orthogonal to the common spin axis, and a twofold rotation (C$_{2y}$) in real space around the Y-axis(shown in Fig.\ref{structure}c), followed by a half-unit-cell translation. The same symmetry classification has been used to discuss NRSS in La$_2$CuO$_4$\cite{vsmejkal2022beyond} and FeSb$_2$\cite{mazin2021prediction}. Further analysis of the symmetry in CrSb$_2$($Pnnm$ space group) reveals the following combination of operations:\\

 $\{E, P, M_z, C_{2z}\} + t\mathcal{T} \{C_{2x}, M_x, C_{2y}, M_y\}$\cite{mazin2021prediction}\\

 Here, the operation $\mathcal{T}$ reverses the spin quantization axis\cite{vsmejkal2022beyond,litvin1974spin,litvin1977spin} and serves as the non-relativistic analog of time-reversal symmetry. The symmetry elements ($E, P, C_{2z}, M_z$) map each magnetic sublattice onto itself and are classified as symmorphic operations. In contrast, the remaining operations, when combined with the translation $t$(half-a-unit cell in the case CrSb$_2$), form nonsymmorphic elements such as glide planes and screw axes. For example\\

 $\mathcal{T}M_xE(k_x,k_y,k_z,\uparrow)=E(-k_x,k_y,k_z,\downarrow)$\\

 For $k_x=0$ (nodal plane), spin up and down bands are degenerate, which corresponds to $\Gamma$(0,0,0) and Z(0,0,1/2) high symmetry points in Fig.\ref{3}(a,b). The presence of such symmetry operation results in alternating spin-splitting along M$^\prime$-$\Gamma$-M [see Fig.\ref{3}(a,b)] and A-Z-A$^\prime$ high symmetry paths.\\

 To investigate the dynamical stability of the relaxed structure, we computed the phonon dispersion curves at 0 and 10 GPa for the M-AFM phase (refer to the supplementary material, Figure S2). The absence of negative frequencies confirms the dynamical stability of the structure. With increasing pressure, a significant blue shift is observed in the phonon spectrum across both the acoustic and optical branches. From the phonon density of states, it is evident that the heavier Sb atoms dominate the low-frequency region, whereas the comparatively lighter Cr atoms contribute predominantly to the high-frequency region.

\section{Conclusion} 
We have performed the first-principles DFT calculation to study the structural, electronic, and magnetic properties of CrSb$_2$ under applied pressure. Our study predicts the structural phase transition from marcasite to HP-II structure at $\sim$11.5 GPa. The experimental results from the temperature-dependent magnetic susceptibility calculation confirm the ground state AFM configuration. The presence of collinear AFM and NRSS of $\sim$0.5eV from the first-principles calculation supports CrSb$_2$ as a strong altermagnet candidate. Interestingly, we have observed a decrease in the NRSS band around the Fermi level and eventually diminishes above $\sim$10 GPa in the HP-II phase, resulting in the flipping of the magnetic characteristic from Altermagnet to conventional AFM. The precise momentum-space positioning of this altermagnetic band splitting holds notable implications for spintronic applications, particularly in enabling the generation of spin-polarized currents in AFMs. Our theoretical findings position CrSb$_2$ as a compelling candidate for experimental exploration within the emerging framework of altermagnetism. Probing techniques such as anomalous Hall conductivity and magneto-optical Kerr effect (MOKE) measurements may offer critical insights into the symmetry-breaking mechanisms, particularly the violation of time-reversal symmetry ($\mathcal{T}$) in this material. 

\section*{Acknowledgments}
RT acknowledges the University Grants Commission (UGC), India, for the Junior Research Fellowship (JRF), ID No. 231620066332.
 
\section*{Author contributions}
\textbf{R. Tamang:} Formal analysis, Visualization, Validation, Literature review, Writing-original draft, writing-review \& editing.\\
\textbf{Shivraj Gurung:} Formal analysis, Visualization, Validation, writing-review \& editing. \\ 
\textbf{Shalika Ram bhandari:} Formal analysis, Visualization, Validation, writing-review \& editing. \\ 
\textbf{Matthew J. Stitz:} Supervision, Formal analysis, Visualization, Validation, writing-review \& editing. \\ 
\textbf{Ganesh Pokharel:}Formal analysis, Visualization, Validation, writing-review \& editing. \\ 
\textbf{Keshav Shrestha}:Formal analysis, Visualization, Validation, writing-review \& editing. \\ 
%		\textbf{Karthik Gopi}: Scripting, Formal analysis, Visualization, Validation, writing-review \& editing. \\
\textbf{D. P. Rai:} Project management, Supervision, Resources, software, Formal analysis, Visualization, Validation, writing-review \& editing. \\ 
	    %\textbf{Samy Brahimi}:Formal analysis, Visualization, Validation, writing-review \& editing.\\
		%\textbf{Samir Lounis}: Supervision, Resources, Formal analysis, Visualization, Validation, writing-review \& editing.
		
		%\begin{acknowledgments}
		%	The research is partially funded by the Ministry of Science and Higher Education of the Russian Federation as part of the World-class Research Center program: Advanced Digital Technologies (contract No. 075-15-2022-312 dated 20.04.2022
		%\end{acknowledgments}
		
	%	\section*{Data Availability Statement}
	%	Data available within the article or its supplementary materials	

%       \nocite{*}		
\bibliographystyle{plain}
\nocite{*}
\bibliography{rsc}
	\end{document}